# Chiral Plasmonic Nanostructures Enabled by Bottom-Up Approaches

Maximilian Urban[1,2,§], Chenqi Shen[3,4,§], Xiang-Tian Kong[5,§], Chenggan Zhu[3], Alexander Govorov[5*], Qiangbin Wang[3*], Mario Hentschel[6*], and Na Liu[1,2*]

[1]*Max Planck Institute for Intelligent Systems, Heisenbergstrasse 3, 70569 Stuttgart, Germany*
[2]*Kirchhoff-Institute for Physics, University of Heidelberg, Im Neuenheimer Feld 227, 69120 Heidelberg, Germany*
[3]*CAS Key Laboratory of Nano-Bio Interface, Division of Nanobiomedicine and i-Lab, Suzhou Institute of Nano-Tech and Nano-Bionics, Chinese Academy of Sciences, Suzhou, 215213 China*
[4]*College of Materials Science and Opto-Electronic Technology, University of Chinese Academy of Sciences, Beijing, 100049 China*
[5]*Department of Physics and Astronomy, Ohio University, Athens, Ohio 45701, USA*
[6]*4th Physics Institute and Research Center SCoPE, University of Stuttgart, Pfaffenwaldring 57, D-70569 Stuttgart, Germany*

[§] These authors contributed to this work equally.
*Corresponding authors. E-mail: govorov@phy.ohiou.edu, qbwang2008@sinano.ac.cn, m.hentschel@pi4.uni-stuttgart.de, na.liu@kip.uni-heidelberg.de

**Abstract**

We present a comprehensive review of recent advances and developments in the field of chiral plasmonics. Chiral plasmonic systems exhibit tailored chiroptical properties vastly outperforming their molecular counterparts. In gaining insight into the working principle, the chiroptical response has become fully tailorable via composition and arrangements of the individual subunits or by shape and size. With advances in micro- and nanofabrication techniques many of these intriguing systems have also been realized in experiment. Here, we focus on bottom-up synthetics methods which do not only allow for rational design and fabrication of structures but most intriguingly in many cases also for post-fabrication manipulation and tuning. Apart from purely plasmonic systems we will also discuss so-called induced chirality, that is, the interaction of chiral molecules with plasmonic excitations which renders achiral plasmonic systems chiral. Chiral plasmonics systems are very promising candidates for investigating conformations and conformational changes on the nanoscale, similarly as in molecular circular dichrosim spectroscopy. The significantly larger signal strength promises few-structure sensitivity. We thus expect the field of chiral plasmonics to attract further widespread applications and attention.

# 1. Introduction

Man is eager to understand the properties of matter and also strives to manipulate matter at will. For centuries, alchemists and scientists have endeavored to control and tailor the properties of matter. A large number of sophisticated techniques have been invented over time, including deposition, etching, doping, synthesis, self-assembly, and many more. One of the most striking material properties of matter is visual appearance, that is, its reflectivity, absorptivity, and most importantly its color. Among the earliest examples of tailored visual appearances is colored glass. Some of the colored glass was later discovered to contain nano-sized gold (Au) particles. Such glass, coined cranberry glass, represents one of the earliest manmade nanostructured materials – obviously unknown at the time.

The origin of the colors lies in the absorption and scattering behavior of the metal nanoparticles in the glass. When light impinges on a metal particle of a few tens of nanometers in diameter, collective oscillations of the quasi-free conduction electrons are excited with respect to the ionic background. One terms this resonant excitation, a plasmon resonance. What makes plasmonic resonances so special? On one hand, the metal particles can be good scatters and absorbers of light. On the other hand, the resonance position, that is, the position of the most efficient scattering and absorption, can be tuned via the size, the material, and the surroundings of the particle. Thus, metal nanoparticles can be tuned to exhibit bright and vivid colors, given their plasmonic resonances are in the visible wavelength range.

What makes plasmonic resonances more interesting is the ability to couple to one another. The harmonic oscillations of the conduction electrons give rise to plasmonic near-fields, which are regions of strong local electromagnetic fields within sub-wavelength volumes. When the local fields of adjacent nanoparticles overlap, coupling among the particles is mediated. In direct analogy to molecular hybridization and the formation of shared orbitals, the plasmonic resonances of adjacent particles can mix and hybridize, giving rise to collective resonant modes spanning the entire structure. The extraordinary number of possible coupling scenarios offers an additional degree of freedom in creating plasmonic nanostructures with interesting optical properties.

In recent years, advances in micro- and nanotechnologies have enabled unprecedented opportunities to manipulate and tailor the optical properties of plasmonic nanostructures according to a designer's wish. Top-down techniques such as electron-beam lithography, focused ion beam etching, direct laser writing, etc., excel in creation of structural arrays and layered interfaces, whereas bottom-up techniques such as self-assembly, chemical synthesis allow for more arbitrary structural geometries and richer biochemical functionalization possibilities. A plethora of complex plasmonic architectures that draw direct analogies to natural molecules have been designed and explored over the years. As a matter of fact, one may argue that significant research on plasmonic nanostructures has taken inspirations from natural molecules due to the beautiful physics and chemistry involved.

One of the most fascinating molecular motifs is chirality. A chiral or handed structure cannot be superimposed with its mirror image. Every chiral object therefore may exist in two distinct handednesses, called two enantiomorphs or enantiomers. Chiral objects are ubiquitous in nature, ranging from macroscopic systems such as seashells to molecular systems such as carbohydrates. Interestingly, many chiral biomolecules are found only in one handedness. For instance, the essential amino acids are all L- enantiomers, meaning the D-enantiomers are absent. This phenomenon is called homochirality, which has intrigued researchers for a long time but a conclusive interpretation remains elusive.

Apart from the structural aspect, chirality can also manifest itself optically. A chiral structure that is optically active can absorb left- and right-handed circularly polarized light (LCP and RCP, respectively) differently. This phenomenon is called circular dichroism (CD). Moreover, a chiral structure can also cause the rotation of linearly polarized light, called optical rotatory dispersion (ORD), resulting from different refractive indices for LCP and RCP light when interacting with the chiral structure. CD and ORD are therefore Kramers-Kronig related.

Chirality is a structural property and closely correlated with the spatial arrangement of the individual constituents in three dimensions. The optical responses of chiral structures are thus widely used in structure-related studies. For instance, CD spectroscopy can be used to reliably report the handedness and even subtle conformational changes of chiral molecules. CD spectroscopy is thus a standard tool in biology and life sciences for chiral discrimination. One major hurdle, however, lies in the weak chiroptical responses of natural chiral molecules.

For most chiral molecules, their CD bands are located in the UV spectral region and the asymmetry factors, which characterize the strength of optical chirality, are very small. Even concentrated solutions result in polarization rotation of light by only a few millidegrees, rendering sensitive detection and investigation of chiral molecules with low amounts very difficult.

The aforementioned findings and challenges have motivated scientists to enhance the strength of molecular chirality as well as to achieve pronounced and tunable chiroptical responses using artificial plasmonic structures. Some studies transfer the intriguing concept of molecular chirality to the realm of plasmonics, whereas others indeed combine the two fields altogether. It has been demonstrated that chiral pure plasmonic structures can exhibit large chiroptical responses, orders of magnitude stronger than those of chiral molecules. These experiments have stimulated further research on the interaction of plasmonic nanoparticles and chiral molecules, with the hope to enhance and possibly to spectrally manipulate optical chirality, rendering the detection of a few chiral molecules possible.

In this article, we will present a comprehensive review of recent advances and developments in the field of chiral nanoplasmonics. We will focus on bottom-up technique related strategies to create chiral plasmonic nanostructures following the outline of (i) plasmon-induced chirality: chiral molecules in the vicinity of achiral plasmonic nanoparticles, (ii) intrinsically chiral colloids: plasmonic particles that are chiral objects themselves, and (iii) plasmonic chirality: achiral plasmonic particles arranged into handed configurations on static and active templates, respectively.

## 2. Plasmon-induced Chirality

2.1 Theoretical Background

CD signals from natural chiral molecules are typically very weak and occur in the ultra-violet spectral range (150 – 300 nm). Considerable efforts have been devoted to enhance CD signals of chiral molecules as well as to bring chiroptical responses to the visible and near-infrared ranges. The local near-fields generated by a metal nanoparticle upon light illumination can mediate the interaction of the particle and a chiral molecule placed in close proximity. The nanoparticle, as a scatterer and absorber of light, can thus be viewed as a reporter of the molecular chirality from the near-field to the far-field. In this case, a new CD line is induced and occurs near the plasmonic resonance position of the metal nanoparticle.

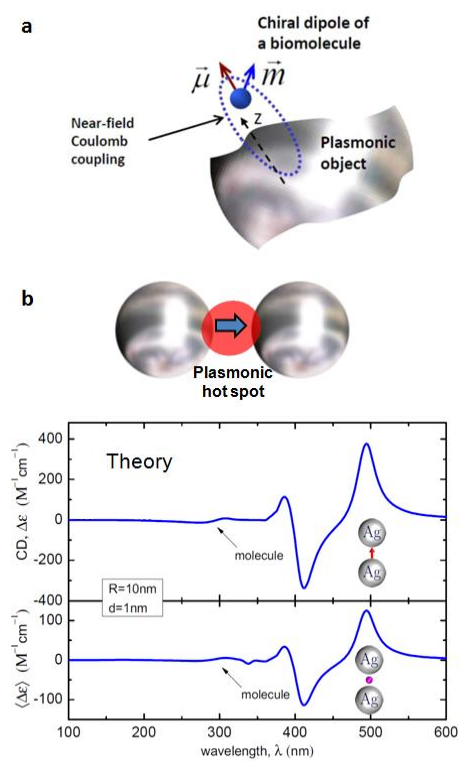

**Figure 1.** (a) Plasmon-induced chirality from a hybrid system containing a chiral molecule in the vicinity of a plasmonic object. Panel *a* adapted from Reference [1] with permission from the American Chemical Society. (b) CD spectra of a Ag dimer with a chiral molecule in the gap region. Upper panel: schematic of the Ag dimer; Middle panel: calculated CD spectrum for the electric dipole of the molecule parallel to the dimer axis; Lower panel: calculated CD spectrum averaged over the molecular orientations. Panel *b* adapted from Reference [2] with permission from the American Physical Society.

Theoretically, the CD responses of a hybrid system containing a chiral molecule and a metal nanoparticle (**Figure 1a**) can be written as follows:

$$\text{CD}_{\text{total}} = \text{CD}_{\text{molecule}} + \text{CD}_{\text{plasmon-induced}} \qquad (1)$$

where $\text{CD}_{\text{molecule}}$ and $\text{CD}_{\text{plasmon-induced}}$ are the CD signals from the differential dissipations of LCP and RCP light inside the molecule and the metal nanoparticle, respectively. The molecular CD modified by the plasmonic near-fields can be written as

$$\text{CD}_{\text{molecule}} = E_0^2 \frac{8}{3}\sqrt{\varepsilon_0}\omega_0 \frac{\Gamma}{|\hbar\omega - \hbar\omega_0 + i\Gamma|^2}\text{Im}\left[\hat{P}\boldsymbol{\mu}_{12}\cdot\mathbf{m}_{21}\right] \quad (2)$$

Here, $E_0$ and $\varepsilon_0$ are the incident electric field and matrix permittivity, respectively. The molecule is described by its electric and magnetic transition dipole moments, denoted by $\boldsymbol{\mu}_{12}$ and $\mathbf{m}_{21}$, respectively. The initial absorption line shape of the molecule is described by the energy, $\hbar\omega_0$, and broadening Γ of the molecular transition. The operator $\hat{P}$ is the matrix of the near-field enhancement at the position of the chiral molecule. The plasmonic near-field enhancement decreases as the distance between the molecule and metal surface increases. In the dipolar limit of the molecule-plasmon interaction, the plasmonic enhancement matrix is proportional to $a^3/d^3$, where $a$ and $d$ are the nanoparticle size and the separation between the molecule and the metal surface, respectively.[4]

The second term in Eq. (1) strongly relies on the shape and size of the metal nanoparticle. It can be explicitly expressed only for simple cases, such as for small spherical nanoparticles with dipolar approximations. In general, this term can be written as

$$\text{CD}_{\text{plasmon}} \propto \text{Im}(\varepsilon_{\text{metal}}) \cdot f_{\text{resonant}} \cdot \text{Im}(\hat{K}\boldsymbol{\mu}_{12}\cdot\mathbf{m}_{21}) \quad (3)$$

Here $\varepsilon_{\text{metal}}$ is the dielectric function of the metal. The operator $\hat{K}$ describes the interactions between the molecule and the plasmonic near-fields. The resonant factor, $f_{\text{resonant}}$, which strongly depends on the size and shape of the nanoparticle, describes the plasmonic enhancement in the metal.

Hotspots generated between two metal nanoparticles can be exploited to amplify the plasmonic field enhancement. Figure 1b shows an example of the plasmon-induced CD from a silver (Ag) dimer. Both the molecular CD lines and the plasmon-induced CD lines are excited. Specifically, compared to the molecular CD line, the plasmon-induced CD responses shift to longer wavelengths and are much stronger. It should be noted that the direction of the electric dipole of the molecule considerably influences the amplitude of the plasmon-induced CD in an anisotropic system. In this case, the strongest plasmon-induced CD occurs, when the molecule dipole $\boldsymbol{\mu}_{12}$ is perpendicular to the metal surfaces. The plasmon-induced CD drops to approximately one quarter of the strongest value, when averaged over the dipole orientations.

The above theory of the plasmon-induced CD can be applied to the cases, in which the metal particles are small compared to the wavelength of light, so that the retardation effects are negligible. However, retardation effects have to be taken into account for large metal structures, which are comparable to the wavelength of light. In this case, the CD response has to be calculated by solving Maxwell's equations. For a chiral medium, Maxwell's equations can be written as

$$\begin{aligned} \mathbf{D} &= \varepsilon \mathbf{E} + i\xi_c \mathbf{B} \\ \mathbf{H} &= \frac{\mathbf{B}}{\mu} + i\xi_c \mathbf{E} \end{aligned} \quad (4)$$

where $\varepsilon = \varepsilon(\omega, \mathbf{r})$ is the permittivity, $\xi_c(\omega, \mathbf{r})$ describes the chiral property of the medium, and $\mu$ is the permeability. For nonmagnetic materials, such as Au, Ag, and chiral biomolecules, we have $\mu = 1$. In general, the CD signal linearly increases with the chiral medium thickness (when the thickness is less than the wavelength of light) for the far-field induction model.

2.2 Experimental Realizations

Attaching chiral molecule ligands on the surfaces of achiral metal nanoparticles has recently become an efficient and convenient route to chemically acquire optical chirality in the visible spectral range [4, 9-14]. In order to achieve strong CD from such complexes, metal particles with different morphologies including nanorods (NRs) [15] and nanocubes [1] have been employed due to their pronounced plasmonic resonances. Chiral molecules such as cysteines [16], peptides [17-19], DNA [20], *etc.*, generally have strong binding affinities to the surfaces of metal particles. The chiroptical responses of the complexes can be precisely tuned through rational control of the chiral ligands on the particle surfaces [21, 22]. For instance, Markovich *et al.* reported Ag nanocubes capped with glutathione, which possessed drastic CD changes upon slightly lowering the pH from 5 to 4.5 as shown in Fig. 2a [5]. A chemical effect of catalyzing the formation of diglutathione took place during the process, resulting in the CD changes. The orientations of the chiral ligands on the particle surfaces can also have a significant impact on the chiroptical properties. Gang *et al.* demonstrated discrete Au/Ag core-shell nanocubes surface-functionalized with DNA [23]. When the DNA ligands were perpendicular to the cube surface, strong plasmon-induced CD was observed. In contrast, when the salt concentration increased, the resulting random DNA orientations led to gradual decrease of the CD responses.

Plasmon-induced chirality can be effectively enhanced when the chiral ligands are in close proximity of the particle surface [13]. In a newly developed strategy, chiral ligands have been embedded within core-shell structures through multi-step metallization processes. For example, Wu *et al.* optimized cysteine-modified Au nanorods via Ag coating [24]. The entrapment of cysteines at the Au-Ag interface amplified the localized electromagnetic fields around the ligands, producing a large enhancement of the CD responses. Wang *et al.* demonstrated a similar approach with Ag shell coating on DNA-capped Au nanorods. In these morphologically diverse core-shell structures, the induced CD was readily manipulated over a broad spectral range by simply controlling the shape anisotropy of the building blocks [25]. Xu *et al.* reported hybrid Au core-DNA-Ag shell nanoparticles with strong and tunable CD using cytosine-rich single-stranded DNA as guiding template for the subsequent Ag shell growth (see Fig. 2b) [6]. In another work [7], the same authors fabricated Au-gap-Ag quasi-spherical nanoarchitectures with interior nanobridged gaps filled with cysteines (see Fig. 2c). The induced CD from the structures was tailored by carefully adjusting the nano-gap distance and the ligand density. Apart from organic and biological chiral molecules, inorganic chiral materials can also couple with metal nanoparticles to form hybrid complexes [26]. Tang *et al.* showed a typical example by coupling chiral mesoporous silica (CMS) with Au nanorods as shown in Fig. 2d [8]. The authors synthesized Au nanorod-CMS core-shell structures with distinct CD signatures in the visible spectral range, enabled by the interactions between the Au nanorods and the chiral molecules in the mesopores of the silica shells.

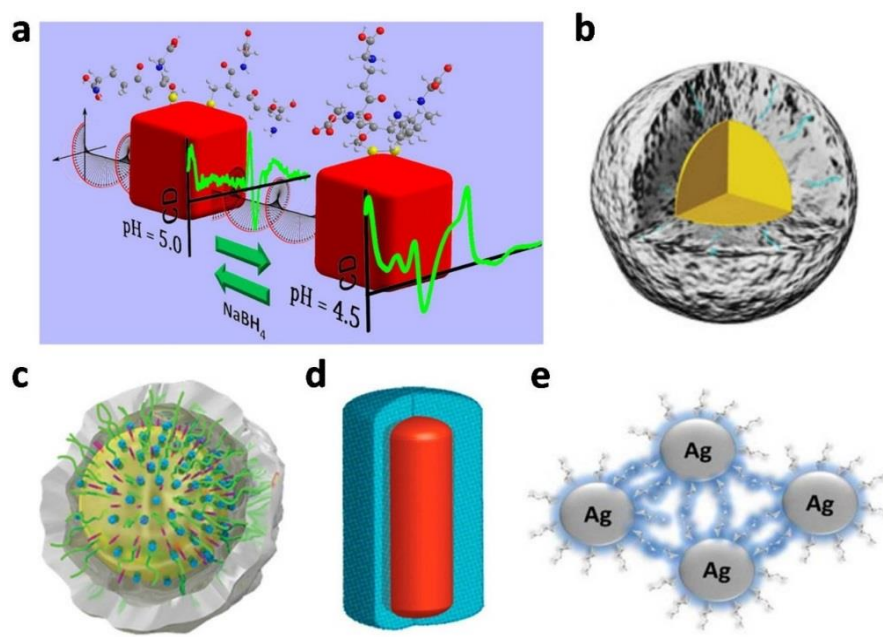

**Figure 2.** Examples of plasmon-induced chirality due to the interaction of chiral molecular species with achiral plasmonic nanoparticles. (a) Ag nanocubes capped with glutathione.[5] (b) Au core-DNA-Ag shell nanoparticles[6]. (c) Au-gap-Ag quasi-spherical structures with nanogaps filled with cysteine.[7] (d) Au nanorod-chiral mesoporous silica core-shell structures.[8] (e) Cholate-coated Ag nanoparticle networks. All images are adapted from the mentioned references with permission from the copyright holder ACS and WILEY-VCH.

Apart from these discrete nanostructures, plasmon-induced chirality has also been achieved in large assemblies, in which nanoparticles were linked together by chiral molecules [27, 28]. One challenge lies in the fact that the metal particle assembly itself may give rise to plasmonic chirality, resulting from the handed arrangements of the metal particles. The CD responses associated with plasmonic chirality in general are much stronger than plasmon-induced CD and also spectrally overlap with the latter [29]. Random nanoparticle aggregates are often used to avoid the generation of plasmonic CD, if only plasmon-induced CD is desired. Tang *et al.* created one-dimensional assemblies of cysteines and Au nanorods[30]. The chiroptical responses were induced by the cysteine molecules attached to the ends of the nanorods. In another example, Markovich *et al.* reported nanostructures composed of crosslinked cholate-coated Ag nanoparticles, illustrated in Figure 1e. In this case, the induced CD was enhanced by localization of the cholate molecules in the large field-strength regions between the nanoparticles [3].

## 3. Intrinsically Chiral Colloids

Chiroptical responses can also be generated from plasmonic nanoparticles, which are intrinsically chiral. The excited plasmonic modes span in the entire chiral object. The CD strength and line shape thus can be readily controlled by tuning the geometry of the particle.

Figure 3a depicts an interesting example studied theoretically. A spherical particle is deformed with a ridge running in a spiral fashion around the particle surface. Similarly, one can create a spiral slit cut into the particle. Such a structure can be called a twister and its counterpart can be called an antitwister. The chiroptical responses arise from the mixing of the particle plasmon modes of different orders due to rather small geometrical distortions. The mode formation is complex and sensitively depends on small geometrical perturbations. In general, the observable chiroptical responses are quite small.

One can increase the chiroptical responses with larger deformations of the particles. In Figure 3b, experimental CD results from plasmonic nanohelices are shown. The nanohelices were fabricated using a sophisticated evaporation scheme that allow for high-quality chiral structures with full control over the handedness and geometry properties. The colloids take the shape of a helix, i.e., a classical chiral object, and the excited plasmonic modes are expected to have a very strong chiral nature. The measured CD spectra underpin this expectation and also show excellent mirror symmetry for left- and right-handed nanohelices (see Figure 3b).

It has been reported that enantioselective control of lattice and shape chirality could be achieved in inorganic nanostructures made of tellurium as shown in Figures 3c. The chiroptical responses of these tellurium nanocrystals were controlled at two hierarchies, atomic lattice and mesoscale shape. Firstly, the tellurium nanocrystals are of a chiral atomic space group, either $P3_121$ or its enantiomorph $P3_221$. Secondly, the overall chiral shapes can be synthesized and controlled with chiral biomolecules, for example with glutathione. Figure 3d shows experimental CD spectra of such nanocrystals. It is assumed that the chiroptical responses mainly originated from the shape of the nanoparticle and the crystal structure played at best a secondary role. The chiroptical responses could for example be further amplified by utilizing the tellurim structures and templates to overgrow them with high quality plasmonic materials.

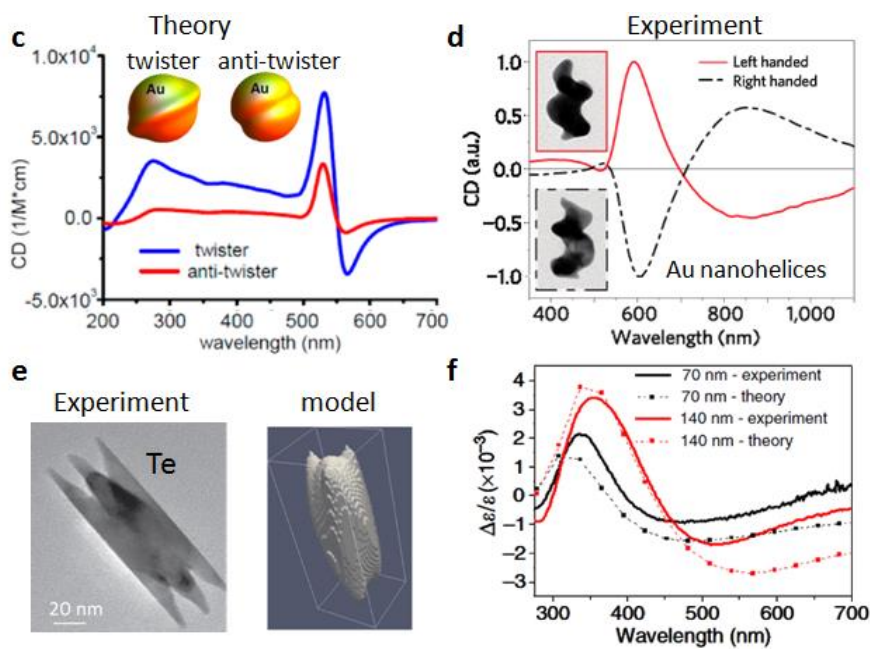

**Figure 4.** Plasmonic structures that are intrinsically chiral. (a) Sketch as well as the theoretical CD response for chiral Au nanocrystals: twister and anti-twister.[14] (b) Experimental CD spectra of solid metal Au nanospirals.[32] Inset: TEM images of the spirals with left (top) and right (bottom) chirality (image dimensions: 85nm × 120nm). (c) A chiral tellurium nanocrystal with chiral atomic symmetry group synthesized with reducing agent (hydrazine) and thiolated chiral biomolecules (glutathione).[33] Left: TEM image; Right: theoretical model. (f) Experimental and theoretical CD spectra of tellurium nanocrystals of the chiral shape shown in (e) and with two different sizes. All images reproduced with permission by the copyright holders Nature publishing group and ACS.

**4. Plasmonic Chirality**

4.1 Theoretical Background

Chiral plasmonic nanostructures composed of individual nanoparticles arranged in chiral geometries can generate strong CD signals in the visible and near-infrared spectral ranges. Figure 4a shows an example of such a system. Spherical and thus achiral nanoparticles are arranged in a helical frame. The plasmonic modes of the individual particles can interact via their respective electromagnetic fields to generate collective modes of the entire structure. Figure 4b shows the characteristic and strong plasmonic CD spectra of such chiral assemblies in the visible spectral range. The dipolar limit applies when the size of the metal nanoparticle $a$ is small compared with the inter-particle distance $R$. In this case, the resulting plasmonic CD follows the following relation

$$\mathrm{CD}_{\text{plasmon-plasmon}} \propto a^{12} / R^{8} \qquad (5)$$

These high power orders clearly show that the strength of the plasmonic CD significantly depends on the nanoparticle size and the distance between them.

Figure 4b depicts the calculated CD spectra of the structures in Figure 4a. Clear bisignate shapes that are well known in CD spectroscopy are observed. Interestingly, the sign of the bisignate shape is a function of the number of particles in the spiral. At the first sight, this is counterintuitive as the handedness of the spirals does not change and thus also the sign should not change. However, as the formation of collective modes in the assembly is strongly influenced by the number and coupling strength between the particles, a change in the number of particles is very likely to cause major spectral shifts in the formed collective modes. This phenomenon is indeed very interesting as the sign change of the bisignate shape is thus not indicative of a change in the handedness of the structure. Here it is related to the strong influence of the plasmonic coupling in the system.

Figure 4c depicts another interesting chiral arrangement. Here, four achiral particles of two different sizes are arranged in an asymmetric chiral frame. [cite Ferry] In contrast to the helices shown in Figure 4a, the handedness of the structure is less obvious and is caused by a shift of the lower smaller particle out of the symmetric position, breaking mirror symmetry. What is interesting in this configuration is the possibility to invert the handedness of the structure by subtle changes. As soon as the particle crosses the mirror plane, the handedness changes and so does the sign of the bisignate shape of the CD response. In this case, the sign change is therefore indeed indicative of a change in structural handedness.

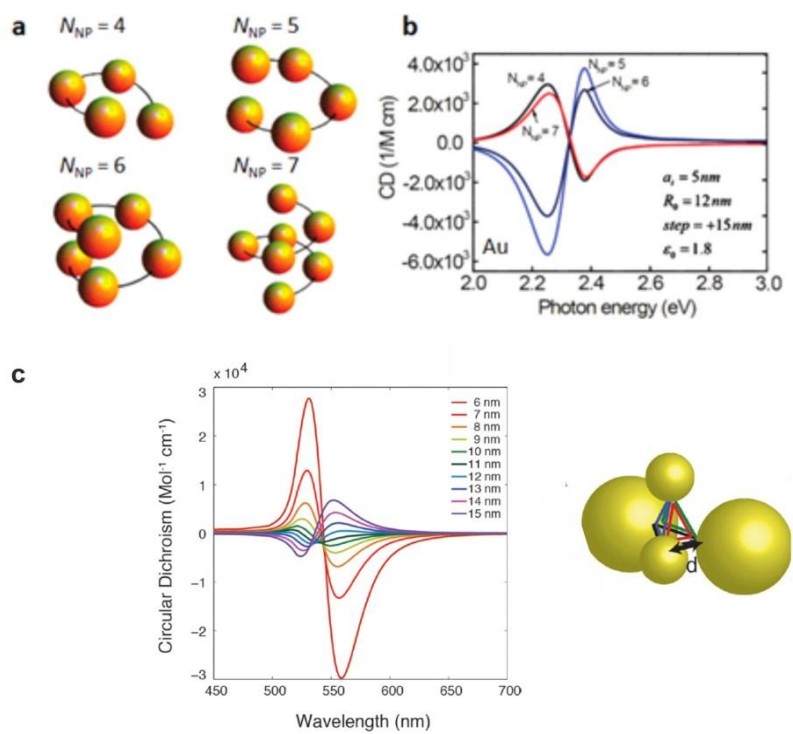

**Figure 4.** Achiral colloids can be grouped such that the resulting arrangement is chiral. Panel (a) depicts the most straightforward design of a spiral. Given the particles are spaces accordingly, the plasmonic modes of the structure can interact with one another generating an overall chiral optical response.[31] Panel (b) depicts the theoretical CD spectra of the assemblies of Au nanoparticles shown in (a) in an aqueous surrounding. Interestingly, the exact shape of the bisignate shape of the CD spectrum sensitively depends on the number of particles. (c) Four achiral particles are arrangement into an asymmetric chiral frame. [cite Ferry]. The handedness is less apparent as for the structures in panel (a). Very small shifts of one of the particles can invert the handedness of the structure and thus the CD response. All images are adapted from the mentioned references with permission from the copyright holder ACS.

(Ferry, V. E., Smith, J. M., & Alivisatos, A. P. (2014). Symmetry Breaking in Tetrahedral Chiral Plasmonic Nanoparticle Assemblies. *ACS Photonics*, *1*(11), 1189–1196)

## 4.2 Experimental Realizations

### 4.2.1 Chiral Templates

Templating is one powerful strategy to organize metal nanoparticles into chiral plasmonic assemblies [39-42]. Useful chiral templates include biomacromolecules [43-45] and polymers [46-49], which can be exploited to pattern a variety of plasmonic building blocks for achieving chiral nanostructures [50-52]. In general, the template-guiding scheme can produce large-scale plasmonic nanostructures with strong chiral responses. In some cases, a contribution due to plasmon-induced chirality from interactions between the chiral template and metal nanoparticles, as discussed in section 2, might also be present [29]. In other words, the resulting CD often possesses a complex origin from both exciton-plasmon and plasmon-plasmon interactions. However, in most cases the overall chiroptical response is dominated by purely plasmonic effects. In the following, we will thus focus on the contribution from plasmonic chirality.

Rosi *et al.* reported peptide-templated chiral assembly of Au nanoparticles and constructed a series of helical superstructures [53-57]. In a first step, rationally designed peptides were assembled to form chain-like helices as frameworks. Au nanoparticles were then synthesized at the Au-binding sites on the peptides by *in situ* reduction. The conjugation sites played a dual role in this methodology: anchoring the Au species for Au nanoparticle nucleation and growth, as well as directing the helical assembly of the as-formed Au nanoparticles. Both LH and RH double-helical arrays with tunable particle sizes were obtained via alteration of the peptide templates and reduction conditions. Strong and tunable CD signals were demonstrated from these superstructures. The experimental procedure was further optimized in a recent work shown in Fig. 5a [34]. Chemical modification of the conjugation sites led to the formation of Au nanoparticle single-helical arrays, which exhibited the highest CD responses among the peptide-directed assemblies. Another template with organogelators has also been widely exploited for *in situ* preparation of chiral plasmonic nanoassemblies [58, 59]. In coordination with noble metal ions such as Au(III) and Ag(I), the designed organogelators were turned into gels through a sol-gel transition process. Au and Ag chiral superstructures and large-scale films templated by the gels were formed by subsequent *in situ* reduction of the ions. In addition, CMS, a more rigid template [60, 61], was utilized for the synthesis of chiral plasmonic materials [62, 63]. Che *et al.* reported Ag nanowire arrays supported by CMS with a distinct multihelix through *in situ* synthesis in the mesopores, which possessed distinct CD responses originated from the interactions between the Ag nanowires [64].

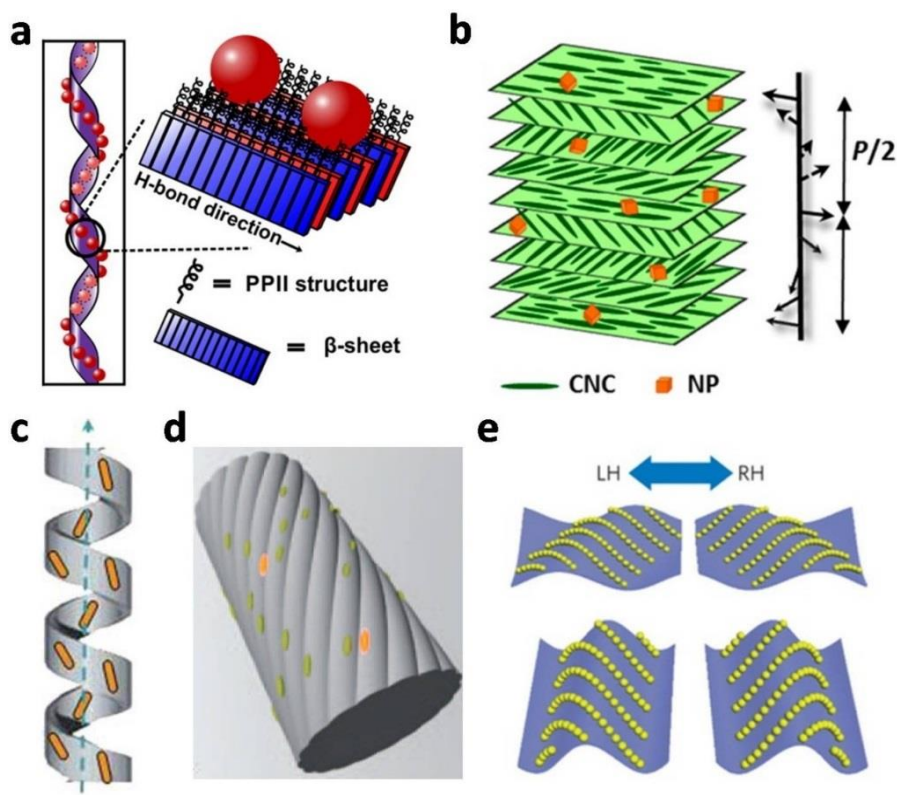

**Figure 5.** Examples of Au nanoparcitles organized with the help of tamplates. (a) Au nanoparticle single-helical arrays templated by peptides.[34] (b) Au nanocubes loaded on cellulose nanocrystals chiral films.[35] (c) Au nanorod/CTAB/lipid chiral hybrid structures.[36] (d) Au nanorod chiral superstructures templated by twisted supramolecular fibers.[37] (e) Chiral nanocomposites with layer-by-layer assembled Au nanoparticle films coated on macroscale twisted PDMS substrates.[38] All images reproduced with permission by the copyright holders ACS, Royal Society of Chemistry, WILEY-VCH, and Nature Publishing Group.

The as-synthesized metal nanoparticles can also be directly absorbed onto the chiral templates. In particular, liquid crystalline polymers offer a broad family of such templates that can be used to efficiently guide the construction of chiral plasmonic assemblies[65-68]. Cellulose nanocrystals are considered as representative materials among them, which show twisted rod-like morphologies [69-71]. For example, a chiral nematic phase with orientational order was formed by aggregation of the anisotropic cellulose nanocrystals. Thin films of these cellulose nanocrystals offer solid orientational host sites for assembling guest nanoparticles [72-75]. After the introduction of plasmonic building blocks, the host-guest composites showed strong CD, much stronger than those from the templates themselves. Xu *et al.* reported Au nanoparticle chiral assemblies loaded on free-standing cellulose nanocrystals films [76]. These host-guest composites not only exhibited tunable and switchable CD responses but also showed unique angle-dependent plasmon resonance properties. Moreover, ultra-long silver nanowires were used as building blocks [77]. The authors realized tunable chiral distribution of the aligned silver nanowires with long-range order through the cellulose nanocrystal liquid crystals mediated realignment. It is noteworthy that chiral organization of anisotropic nanoparticles such as nanowires and nanorods usually display more pronounced CD responses when compared with the spherical counterparts owing to their larger resonance dipole moments. Kumacheva *et al.* demonstrated a similar system, in which Au nanorods were loaded on cellulose nanocrystal films to generate chiral plasmonic complexes [78]. Later, the authors tuned the chiroptical properties using Au nanorods with different aspect ratios and different types of cellulose nanocrystals as shown in Fig. 5b [35]. Nanocubes containing octahedral Au cores and Ag shells were integrated in these chiral films as well. Apart from cellulose nanocrystals, other cholesteric liquid crystals were also utilized as chiral templates. Chung *et al.* reported a Au nanorod/CTAB/lipid, a chiral hybrid superstructure, in which the Au nanorods were well dispersed on a CTAB/lipid supramolecular template (see Fig. 5c) [36]. Strong chiral responses were achieved due to the helical alignment of Au nanorods in the CTAB/lipid nanoribbons. Similar to a molecular liquid crystalline phase, helical geometries of the Au nanorod superstructures and the resulting CD responses were found to show interesting temperature dependence.

Expanding the scale of the plasmonic assemblies has proved to be an efficient approach for acquiring stronger CD as more intense coupling effects can take place between the numerous metal nanoparticles. As shown in Fig. 5d, Liz-Marzán *et al.* demonstrated Au nanorod superstructures templated by chiral fibers [37]. Specifically, Au nanorods were adsorbed onto the supramolecular fiber scaffolds with twisted morphologies through particular non-covalent interactions to form 3D helical ordering. These remarkable composites possessed lengths up to micrometers. The Au nanorods on the fibers were strongly coupled due to their nearly end-to-end arrangement. As a result, giant chiral responses as well as unprecedented levels of anisotropy factors in the visible to near infrared regime were observed from the superstructures. Additionally, the CD signals and the anisotropy factors of the composites could be precisely tailored via tuning of the Au nanorod concentration. In a recent work, Kotov *et al.* realized chiral plasmonic composites using macroscale templates (see Fig. 4e) [38]. Exploited as templates, macroscale elastic poly(dimethylsiloxane) (PDMS) substrates were conformally coated with layer-by-layer assembled Au nanoparticle films. The substrates were mechanically twisted in opposite directions to form LH and RH structures, respectively. The chiral responses could be reversibly reconfigured and cyclically modulated through mechanical stretching. The layered composites of Au nanoparticles exhibited strong CD observed as distinct features around the plasmonic resonance positions.

4.2.2 DNA Origami Templates

Among a variety of materials for self-assembly, DNA represents one of the most attractive candidates largely owing to its unprecedented programmability and sequence specificity. Structural DNA nanotechnology takes the unique advantage of highly sequence-specific interactions between complementary DNA strands to create DNA nanostructures that can be programmed in a rational manner. The breakthrough in structural DNA nanotechnology came with a concept called DNA origami, which involves the folding of a long DNA scaffold strand by hundreds of designed short oligonucleotides [83]. The so-called staple oligonucleotides are hybridized to the DNA scaffold through Watson-Crick base pairing to create arbitrary 2D or 3D nanostructures. As the positions of the staples within the origami are predetermined, the formed DNA nanostructure is fully addressable. Functionalization of DNA origami is enabled by capture strands that extend from the origami. This allows for DNA origami-assembly of metal nanoparticles, quantum dots, fluorescent dyes, and other entities. Compared to other assembly strategies based on discrete DNA strands, the DNA origami approach offers several major advantages: The folding is more robust against imperfections in the stoichiometry of the individual DNA strands and the formed origami template is more rigid, enabling nanoparticle assemblies with defined configurations.

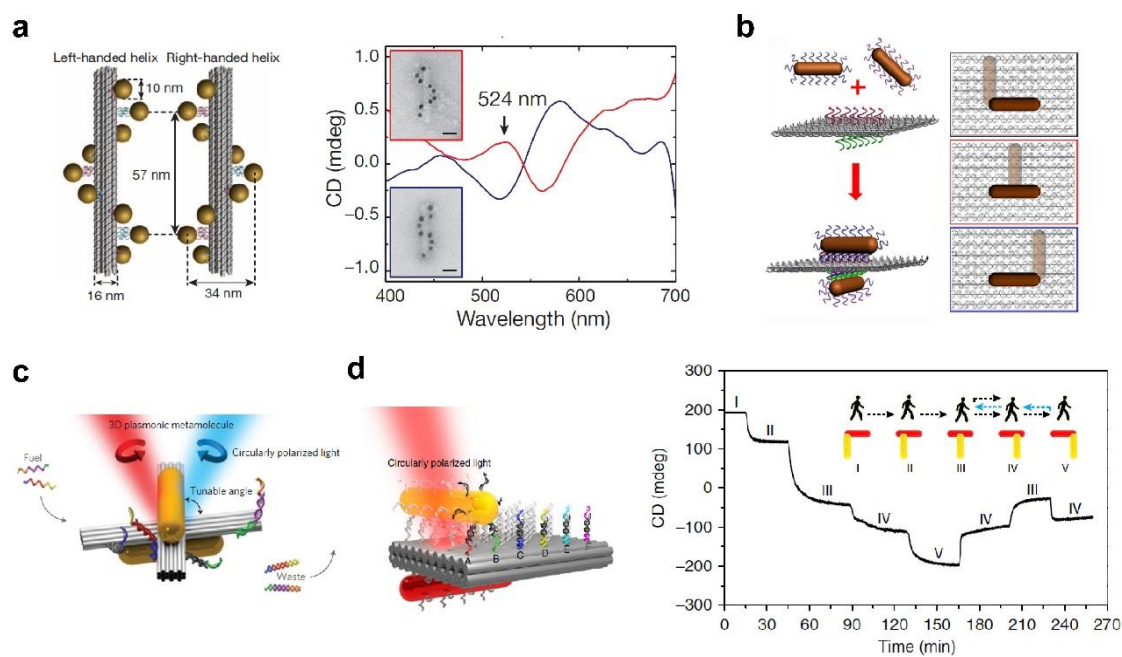

**Figure 6.** (a) (left) Schematic of nanoparticle helices, (right) Experimental results from CD spectroscopy and TEM.[79] The left- and right-handes helices show a clear and mirror symmetric CD response in the visible wavelength range. (b) Different arrangements of Au nanorod dimers assembled on a rectangular DNA origami structure.[80] The assembly strategy allows to straightforwardly create left- and right-handed as well as achiral structures. (c) Schematic of a 3D reconfigurable plasmonic metamolecule consisting of two nanorods on a DNA template. The relative orientation (i.e., the rotation angle) can be controlled post-assembly and switched repeatably.[81] (d) (left) Schematic of a plasmonic Au nanorod walker assembled on a rectangular DNA origami. The lower rod is immobilized, while the upper one can walk over the origami sheet by addition of the appropriate DNA strands to the solution. (right) In situ monitoring of the walking process. The numbers I-IV represent different states of the system (see inset) and clearly demonstrate the stability and reproducibility of subsequent walking-stepc.[82] Figures adapted with permission from the Nature Publishing Group and ACS.

Yan and coworkers were the first to assemble multiple Au nanoparticles on DNA origami[84], which has inspired researchers to step beyond simple plasmonic geometries such as dimers and trimers. In 2012, the experimental realization of chiral plasmonic nanostructures was demonstrated by Kuzyk *et al.* [79]. The authors created Au nanoparticle helices that exhibited designated chiroptical properties. The Au nanoparticles were wound around a rigid DNA origami core in both handedness (see Fig. 6a). The helical arrangementsof Au nanoparticles gives rise to a characteristic bisignate CD spectrum centered on the plasmon resonance frequency of the individual nanoparticles. As expected, Au nanoparticle helices of opposite handedness showed mirrored CD signals. Shen *et al*. as well demonstrated the formation of Au nanoparticle helices using DNA origami [85]. A 2D rectangular DNA origami was first utilized to organize Au nanoparticles precisely at well-defined binding sites along two linear chains. 3D Au nanoparticle helices were then achieved by rolling and stapling the 2D origami sheets into tubes.

The essential ingredients for achieving strong plasmonic chirality were studied in detail by Shen *et al* [86]. The authors utilized a rigid addressable DNA origami template to precisely organize four nominally identical Au nanoparticles into a 3D asymmetric tetramer. It was demonstrated that the structural asymmetry and the plasmonic resonant coupling were both required for achieving strong chiral responses. DNA origami also enabled chiroptical responses from centrosymmetric octahedral DNA-frames[87]. In this case, the Au nanoparticles were arranged at prescribed locations on the vertices of the frame. The CD responses resulted from the use of different-sized nanoparticles arranged in an octahedral herocluster.  Later, the assembly of multiple chiral building blocks into larger chiral architectures has also been investigated[88]. Urban *et al*. demonstrated hierarchical assembly of plasmonic toroidal metamolecules, consisting of four identical origami-templated helical building blocks. Plasmonic toroidal metamolecules showed a stronger chiroptical response than the constituent building blocks including helical monomers and dimers.

Compared to spherical nanoparticles, anisotropic Au nanrods possess stronger resonant diploe moments and more efficient spectral tunability. In addition, it is possible to create a 3D chiral nanostructure using only two identical Au nanrods. Nevertheless, assembly of Au nanrods on DNA origami with a high yield is considerably more challenging than assembly of isotropic nanoparticles, which are not orientation dependent. Lan *et al*. [80] reported Au nanrod crosses with controlled interparticle distance and spatial orientations on rectangular bifacial DNA origami (see Fig 4 b). Tuning the docking positions of the Au nanrods allowed for the realization of left- and right-handed nanostructures. In a subsequent work, the authors demonstrated helical Au nanrod superstructures with tailored chirality in a programmable manner[89]. By designing the arrangement pattern of the DNA capture strands on both sides of a 2D DNA origami template, Au nanrods were assembled into Au nanrod helices with intercalated origami. Left- and right-handed Au nanrod helices were fabricated by tuning the mirrored-symmetric patterns of the capture strands on the origami.

Efforts have been made to control the position and orientation of single stranded DNA on Au nanoparticles [90]. A recent work by Zhang *et al*. [91] combined the advantage of the DNA origami technique and the assembly strategy of discrete DNA strands. By attaching a Au nanoparticle to a set of patterned DNA strands on origami, the positions of the strands on the Au nanoparticle could be controlled. Subsequent detachment of the modified Au nanoparticle allowed for assembly with additional Au nanoparticles to form chiral plasmonic nanostructures. The authors demonstrated great control over the chirality of pyramidal nanoparticle groupings.

The most unique feature of the DNA origami technique lies in the possibility to reconfigure assembled nanostructures. This allows for the generation of dynamic optical elements. When designing dynamic nanoscale devices, three prerequisites are of paramount importance. First, an efficient energy source for triggering conformation changes at the nanoscale is crucial. Equally important is the reversible control over conformation of individual nanostructures. Last but not least is the ability to report such nanoscale conformation changes and translate them into tunable functionalities. Kuzyk *et al*. [81] demonstrated the first reconfigurable 3D plasmonic nanostructures, which executed DNA-regulated conformational changes at the nanoscale (see Fig. 6c). DNA served as both a construction material to organize AuNRs in three dimensions, as well as fuel for driving the nanostructure to distinct conformational states through toehold-mediated strand displacement reactions. Simultaneously, the nanostructure worked as its own optical reporter, transducing the conformational changes into CD changes in real time. Two AuNRs were hosted on a reconfigurable DNA origami template. By adding specifically designed DNA fuel strands, the plasmonic nanostructures could be switched between different conformational states characterized by distinct CD spectra. The same authors extended the concept to a light switchable DNA lock where a photoresponsive active site was introduced. [92] In a recent work, Kuzyk *et al*. reported the realization of DNA-assembled reconfigurable plasmonic nanostructures, which could respond to a wide range of pH changes [93]. Such programmability allowed for selective reconfiguration of different plasmonic species coexisting in solution through pH tuning by utilizing pH-sensitive DNA locks as active sites to trigger structural regulation of the chiral plasmonic nanostructures over a wide pH range.

In a study by Zhou *et al*. [82], the authors demonstrated a dynamic plasmonic system, in which a Au nanorod could perform stepwise walking directionally and progressively on DNA origami (see Fig. 6 d). The nanoscale steps were in situ monitored by CD spectroscopy. The key idea was to create a plasmonically coupled system, in which a walker and a stator, constituted a conformationally sensitive geometry. When the walker carried out stepwise movements, it triggered a immediate spectral response that could be read out optically. As a result, stepwise walking with step size far below the optical resolution limit could be optically discriminated in real time. Urban *et al*. further modified the system and showed a plasmonic walker couple system, in which two nanorod walkers could independently or simultaneously perform stepwise walking powered by DNA hybridization along the same DNA origami track [94].

## 5. Summary and Outlook

We have reviewed the current trends and directions in chiral nnaoplasmonics. Due to the wealth of research, it is not possible to give credit to all the interesting work conducted, in particular as this review has focused on bottom-up technique related strategies.

Firstly, chiral molecules, which are surprisingly abundant in nature, can interact with achiral plasmonic structures and render the overall composite chiral and lead to chiroptical responses. While the observed signal strength is comparably small, the effect is interesting for several reasons. On one hand, the interaction between entities with remarkably different size scales and dipole strengths is per se interesting, as partially unexpected. On the other hand, the induced CD from the composite is generally observable centered at the resonance positions of the plasmonic nanoparticles, which can effectively shift the observation window for molecular CD from the UV towards the visible region. Moreover, researchers have the hope to achieve stronger chiral effects from the composites than from pure chiral molecules.

Secondly, we discussed in detail how to use achiral plasmonic building blocks and assemble them into handed geometries. Templates, such as peptides or PDMS, can be used to organize nanoparticles of different sizes and shapes with high efficiency. These assemblies of nanoparticles normally take the structure of the underlying template. While the control over structurally parameters is comparably limited, the technique can be straightforwardly scaled to large quantities. The DNA self-assembly techniques offer tremendous control over the geometry and placement of nanoparticles. A long DNA strand is folded with the help of stable strands into nearly arbitrary structures, depending on the design of the DNA strands. These origami templates, which can for example take the shape of a sheet, can be dressed with additional capture strands that extend from pre-programmed position of the template. With these strands nanoparticles can be selectively placed on the origami with excellent yield. What is even more fascinating is the possibility to manipulate the structures after assembly by appropriate design of the capture sides. So called fuel strands can unbind particles and rebind them to other position, changing the conformation of the molecules. While all these aspects are very intriguing, the scalability of the technique is limited at this time and design and fabrication are challenging and costly.

Lastly, there is also the possibility to create intrinsically chiral colloids. In these cases the particles take the shape of helices or similar handed structures. Depending on the degree of asymmetric, the chiroptial response can be comparably small. However, true solid metal helices have shown huge CD responses, even in the visible wavelength range.

A significant portion of the reported research has been motivated by the importance of chirality and CD spectroscopy in general and the overall fascination of the concept of chirality. One of the most striking properties of chirality is its unique ability to report three-dimensional information. Chirality is in fact a truly three-dimensional property. CD spectroscopy has consequently extensively been used to analyze molecular conformation and observe conformational changes. There is still large interest in utilizing plasmonic structures for enhanced CD spectroscopy or for conformational analysis. This strategy is mostly motivated by the very efficient light scattering and absorption of plasmonic structures which largely outperforms molecular systems. Whether or not this goal can be reached and to what degree it can be exploited remains to be seen. However, the research conducted and reported here has unambiguously shown that chiral structures are uniquely suited to report nanoscale arrangements in form of far-field scattering, as for example in case of the plamsmonic walker.[82] The field of chiral plasmonics therefore remains interesting and many fascinating problems lie ahead to be studied.

DISCLOSURE STATEMENT

The authors are not aware of any affiliations, memberships, funding, or financial holdings that might be perceived as affecting the objectivity of this review.

ACKNOWLEDGMENTS